\documentclass[12pt]{article}
\usepackage{epsfig}
\usepackage{psfrag}
\usepackage{latexsym}
\usepackage{amssymb}
\usepackage{amsmath}
\usepackage{mathrsfs}

\def\beq{\begin{equation}}
\def\eeq{\end{equation}}
\def\bea{\begin{eqnarray}}
\def\eea{\end{eqnarray}}
\def\bet{\begin{tabular}}
\def\eet{\end{tabular}}
\def\bes{\begin{subequations}\bea}
\def\ees{\eea\end{subequations}}

\newcommand{\ba}{\begin{array}{c}}
\newcommand{\baz}{\begin{array}{cc}}
\newcommand{\bad}{\begin{array}{ccc}}
\newcommand{\ea}{\end{array}}

\newcommand{\num}{\nu_{\mu}}
\newcommand{\nut}{\nu_{\tau}}

\def\om{\omega}

\def\be{\begin{equation}}
\def\ee{\end{equation}}
\def\bc{\begin{center}}
\def\ec{\end{center}}
\def\bea{\begin{eqnarray}}
\def\eea{\end{eqnarray}}

\def\nn{\nonumber}

\catcode`@=11
\def\marginnote#1{}
\newcount\hour
\newcount\minute
\newtoks\amorpm
\hour=\time\divide\hour by60
\minute=\time{\multiply\hour by60 \global\advance\minute by-\hour}
\edef\standardtime{{\ifnum\hour<12 \global\amorpm={am}%
        \else\global\amorpm={pm}\advance\hour by-12 \fi
        \ifnum\hour=0 \hour=12 \fi
        \number\hour:\ifnum\minute<10 0\fi\number\minute\the\amorpm}}
\edef\militarytime{\number\hour:\ifnum\minute<10 0\fi\number\minute}
\def\draftlabel#1{{\@bsphack\if@filesw {\let\thepage\relax
   \xdef\@gtempa{\write\@auxout{\string
      \newlabel{#1}{{\@currentlabel}{\thepage}}}}}\@gtempa
   \if@nobreak \ifvmode\nobreak\fi\fi\fi\@esphack}
        \gdef\@eqnlabel{#1}}
\def\@eqnlabel{}
\def\@vacuum{}
\def\draftmarginnote#1{\marginpar{\raggedright\scriptsize\tt#1}}
\def\draft{\oddsidemargin 0.0truein
        \def\@oddfoot{\sl preliminary draft \hfil
        \rm\thepage\hfil\sl\today\quad\militarytime}
        \let\@evenfoot\@oddfoot \overfullrule 3pt
        \let\label=\draftlabel
        \let\marginnote=\draftmarginnote
   \def\@eqnnum{(\theequation)\rlap{\kern\marginparsep\tt\@eqnlabel}%
\global\let\@eqnlabel\@vacuum}  }
\catcode`@=12
\begin{document}
\begin{titlepage}
\vspace*{-1cm}
\phantom{hep-ph/***} 
\hfill{DFPD-07/TH/19}


\vskip 2.5cm
\begin{center}
{\Large\bf Fermion Mass Hierarchies and Flavour Mixing\\
\vskip .2cm
from a Minimal Discrete Symmetry}
\end{center}
\vskip 0.2  cm
\vskip 0.5  cm
\begin{center}

{\large Ferruccio Feruglio}~\footnote{e-mail address: feruglio@pd.infn.it} and 
{\large Yin Lin}~\footnote{e-mail address: Lin@pd.infn.it}
\\
\vskip .1cm
Dipartimento di Fisica `G.~Galilei', Universit\`a di Padova 
\\ 
INFN, Sezione di Padova, Via Marzolo~8, I-35131 Padua, Italy
\\
\end{center}
\vskip 0.7cm
\begin{abstract}
\noindent
We construct a simple model of fermion masses based on a spontaneously broken $S_3\times Z_3$ flavour group.
At the leading order, in the neutrino sector $S_3$ is broken down to a $\nu_\mu-\nu_\tau$ parity subgroup that enforces
a maximal atmospheric mixing angle and a vanishing $\theta_{13}$. In the charged lepton sector the $\nu_\mu-\nu_\tau$ parity
is maximally broken and the resulting mass matrix is nearly diagonal. The charged lepton mass hierarchy is
automatically reproduced by the $S_3$ symmetry breaking parameter alone. A careful analysis shows that,
after the inclusion of all relevant subleading effects, the model predicts $\theta_{23}=\pi/4+O(\lambda_c^2)$ 
and $\theta_{13}=O(\lambda_c^2)$, $\lambda_c$ denoting the Cabibbo angle. A simple extension to the quark sector is also illustrated,
where the mass spectrum and the mixing angles are naturally reproduced, with the exception of the mixing angle between the first two
generations, that requires a small accidental enhancement.
\end{abstract}
\end{titlepage}
\setcounter{footnote}{0}
\vskip2truecm
%
\section{Introduction}
On the eve of the tenth anniversary of the SuperKamiokande (SK) data on atmospheric neutrinos, that shook the whole field
and gave rise to a decade of incredible excitement and activity, neutrino oscillation parameters are known to a 
sufficiently high precision to make desirable a theoretical description going beyond the mere fitting procedure.
In particular the leptonic mixing pattern, so different from the one in the quark sector, provides a 
non-trivial theoretical challenge. The present data \cite{data}:
\be
\theta_{12}=(34.5\pm 1.4)^0~~~,~~~~~~~\theta_{23}=(42.3^{+5.1}_{-3.3})^0~~~,~~~~~~~\theta_{13}=(0.0^{+7.9}_{-0.0})^0~~~,
\label{angles}
\ee
are fully compatible with the so-called tri-bimaximal (TB) mixing pattern, where
\be
\sin^2\theta_{12}=\frac{1}{3}~~~(\theta_{12}=35.3^0)~~~,~~~~~~~\sin^2\theta_{23}=\frac{1}{2}~~~,~~~~~~~\sin^2\theta_{13}=0~~~.
\ee
Several theoretical mechanisms leading to a nearly TB mixing have been suggested in the last years \cite{TB3, TB1, TB2, altarelli}.
The TB mixing has the advantage of correctly describing the solar mixing angle, which, at present, is the most precisely
known. Indeed, its 1$\sigma$ error, $1.4$ degrees corresponds to less than $\lambda_c^2$ radians, where $\lambda_c\approx 0.22$ 
denotes the Cabibbo angle. 

Reaching a similar sensitivity on $\theta_{23}$ and $\theta_{13}$ will require some more years
of work, but it is a remarkable feature that the central values of these angles remained surprisingly stable in the last ten years.
The value of $\theta_{23}$ quoted above is largely dominated by the SK data. It is however noticeable that the 
independent determinations of $\theta_{23}$ by MACRO \cite{macro}, K2K \cite{k2k}, MINOS \cite{minos} and SK \cite{sk},
analyzed in a two-flavour framework, all select $\sin^22\theta_{23}=1$ as best fit value. By removing the 
boundary $\sin^22\theta_{23}\le 1$, K2K, MINOS and SK prefer $\sin^22\theta_{23}$ slightly outside the physical region.
Notice that the value of $\theta_{23}$ extracted from three-flavour global fits, quoted in eq. (\ref{angles}), is slightly non-maximal.
Such an effect becomes manifest when we move from the two-flavour analysis to the three-flavour one \cite{nonmax}. In the last case,
the dependence on $\Delta m^2_{12}$ is included in the analysis of atmospheric neutrino data and the deviation
from maximality is due to the presence of a small excess of events in the sub-GeV electron sample of SK, 
which the two-flavour analysis cannot completely account for. At the moment such a deviation is not statistically 
significant, but future, more precise data might confirm that the maximality of $\theta_{23}$ is violated 
at the $\lambda_c^2$ level.
The upper bound on $\theta_{13}$ is dominated by the CHOOZ data \cite{chooz}, though the preference for a small
$\theta_{13}$ is also present in the solar and the atmospheric data samples. Moreover, the recent
significant error reduction on $|\Delta m^2_{23}|$ by MINOS has also sharpened the CHOOZ bound on $\theta_{13}$,
which depends on  $|\Delta m^2_{23}|$. From the theory point of view, maximal and vanishing mixing angles are
special and many theoretical efforts have been devoted to explain how $\theta_{23}=\pi/4$ and/or $\theta_{13}=0$
can be generated. 

As a matter of fact, in the so-called flavour basis, the most general neutrino mass matrix
giving rise to $\theta_{23}=\pi/4$ and $\theta_{13}=0$ displays a $\nu_\mu-\nu_\tau$ parity
symmetry \cite{numunutau}. Such a parity symmetry, when extended to the whole theory, is expected to be 
broken in the charged lepton sector by the large hierarchy $m_\mu/m_\tau\ll 1$.
In a realistic model 
the breaking effects, responsible for non-vanishing $\theta_{23}-\pi/4$ and
$\theta_{13}$, should have only a small impact on the neutrino sector. Assuming diagonal
charged leptons, textures supported by the $\nu_{\mu}-\nu_{\tau}$ symmetry,
including breaking effects, have been widely studied in the literature \footnote{The
exchange symmetry between the second and the third generations has also been 
adopted as a texture symmetry not only for neutrinos but for all fermions \cite{others2}.
Both lepton and quark mixing angles can be reproduced and there are enough parameters to
fit the fermion masses, without however explaining their hierarchies.} 
\cite{breaking0, breaking1, breaking2}.
Special attention have been paid to the effect of small $\nu_{\mu}-\nu_{\tau}$ symmetry breaking terms 
coming from the neutrino sector on the values $\cos2 \theta_{23}$ and $\sin^2 \theta_{13}$.
Depending on the type of neutrino mass hierarchy, correlations among small quantities such as 
$\sin^2 \theta_{13}$, $\cos2 \theta_{23}$ and $R=\Delta m^2_{sol} /\Delta m^2_{atm}$ 
have been investigated. In particular, in simple frameworks, it turns out that $\sin^2 \theta_{13}$ 
is strongly suppressed by $R$ for normal hierarchy \cite{breaking0}.
Since the Dirac phase is absent in the $\nu_{\mu}-\nu_{\tau}$ symmetric limit,
we expect that deviations of $\theta_{23}-\pi/4$ and $\theta_{13}$ from zero 
are sensitive to CP phases (including Majorana phases) \cite{breaking0}.
Moreover, the breaking of the $\nu_{\mu}-\nu_{\tau}$ symmetry can be driven entirely by the introduction
of CP violating phases \cite{breaking2}. In this case strict correlations between the CP violation
and the broken $\nu_{\mu}-\nu_{\tau}$ symmetry can be established
\footnote{CP violating phases associated to $\nu_{\mu}-\nu_{\tau}$ breaking effects are also
relevant in leptogenesis. In simple realizations of the seesaw mechanism, an exact 
$\nu_\mu-\nu_\tau$ symmetry in the Dirac mass matrix and in the right-handed neutrino mass matrix
implies a vanishing primordial lepton asymmetry. A successful leptogenesis 
requires appropriate extensions of this scheme \cite{lepto}.}.
However, in a general framework  where $U_l$, the contribution to lepton mixing from the charged leptons, 
is only approximately equal to the unit matrix, many of these correlations are relaxed.
For this reason, it is particularly instructive to explore models of fermion masses 
providing a consistent and unified picture of the neutrino sector and of the hierarchy 
among charged lepton masses. Only in such a context it will be possible to keep under control
all possible breaking effects of the $\nu_{\mu}-\nu_{\tau}$ symmetry and to achieve a 
realistic next-to-leading order prediction for $\cos2 \theta_{23}$ and $\sin^2 \theta_{13}$.

Nowadays promising candidates for a unified picture of fermion mass hierarchies and flavour mixing
are the models based on spontaneously broken flavour symmetries. The mixing angles, in particular the
leptonic ones, are best understood by a mechanism of vacuum misalignment occurring in theories with non-abelian
flavour symmetries \footnote{Spontaneously broken discrete symmetries with preserved sub-groups 
can play also an important role in explaining the Cabibbo angle, see for instance \cite{Cabibbo}.}. 
In the various fermion sectors of the theory (up quarks, down quarks, charged leptons and neutrinos)
the symmetry is broken along different directions in flavour space and the corresponding diagonalizations require
misaligned unitary transformations, which end up in the desired mixing pattern.
Such breaking schemes are easy to arrange in SUSY models based on small discrete symmetry groups,
where the discussion of vacuum alignment is particularly simple and transparent. Also
the fermion mass hierarchies can be achieved via spontaneous breaking of the flavour symmetry.
However, in most cases, a separate component of the flavour group is exploited to this purpose.
Quite frequently the flavour group is of the type $D\times U(1)_{FN}$ where
$D$ is a discrete component that controls the mixing angles and $U(1)_{FN}$ is an abelian continuous symmetry
that describes the mass hierarchy, along the lines of the original Froggatt-Nielsen proposal \cite{FN}.
It would be desirable to have a more economical model where the same flavon fields
producing the mixing pattern via VEV misalignment are also responsible for the mass hierarchies.
Models of this type based on the gauged flavour groups $SU(3)$ and
$SO(3)$ exist in the literature \cite{su3}. In these models the charged fermion mass hierarchies are obtained
via a flamboyant flavour symmetry breaking sector and a particular choice of the messenger
scales. It would be interesting to identify a kind of minimal flavour group with few flavon fields, able to provide a decent description of the main features
of the fermion mass spectrum, including the approximate vanishing of $\theta_{13}$ and of $\theta_{23}-\pi/4$.

In this paper we illustrate a model for lepton masses based on the small discrete group 
$S_3\times Z_3$. The non-abelian factor $S_3$ is spontaneously broken by a special vacuum misalignment.
This guarantees the relations $\theta_{23}=\pi/4$ and $\theta_{13}=0$
at the lowest order in the expansion parameters $\langle\varphi\rangle/\Lambda\ll 1$ describing the symmetry breaking
of $S_3$. The same expansion also provides the required suppression factors that
``explain'' the observed hierarchies among charged lepton masses. There is no need of additional flavons
or of additional group factors to describe, at the level of orders of magnitude, all lepton masses.
The $Z_3$ factor remains unbroken to forbid  unwanted couplings between different fermion sectors. 
It is also of great experimental interest to establish at 
which order in $\langle\varphi\rangle/\Lambda\ll 1$ the leading order 
results for $\theta_{23}$ and $\theta_{13}$ are potentially violated. After a detail analysis of all breaking effects, we get:
\be
\theta_{23}=\frac{\pi}{4}+O\left(\lambda_c^2\right)~~~,
~~~~~~~\theta_{13}=O\left(\lambda_c^2\right)~~~,
\ee
where $\lambda_c$ is the Cabibbo angle. The structure of the model is very simple. 
The main feature is the spontaneous breaking of $S_3$ 
by two sets of flavon fields, $\varphi_e$ and $(\varphi_\nu,\xi)$. Thanks to a vacuum alignment property,
$(\varphi_\nu,\xi)$ breaks $S_3$ down to a $\nu_\mu-\nu_\tau$ parity subgroup, while $\varphi_e$
breaks $S_3$ completely. In particular the $\nu_\mu-\nu_\tau$ parity is maximally broken by $\varphi_e$.
These two breaking patterns are selectively supplied to the Yukawa couplings of the theory. At the leading order of
the expansion in $VEV/\Lambda$, $\Lambda$ being the cutoff of the theory, $\varphi_e$ affects only the charged lepton mass matrix,
which is nearly diagonal, while $(\varphi_\nu,\xi)$ controls the neutrino mass matrix, which remains invariant under
the $\nu_\mu-\nu_\tau$ exchange, thus guaranteeing $\theta_{23}=\pi/4$ and $\theta_{13}=0$.
Such a scheme is particularly economic in terms of new fields and new symmetries.
A distinguished feature of our model is that the hierarchy of the masses in the charged lepton sector is
also controlled by the $S_3$ breaking and $m_\tau$, $m_\mu$, $m_e$ get their first non-vanishing contribution
at the order $\langle \varphi_e\rangle/\Lambda$, $(\langle \varphi_e\rangle/\Lambda)^2$, $(\langle \varphi_e\rangle/\Lambda)^3$,
respectively. This allows to estimate $\langle \varphi_e\rangle/\Lambda\approx\lambda_c^2$.
We also analyze all the effects that can modify $\theta_{23}=\pi/4$ and $\theta_{13}=0$. We discuss the sub-leading operators
and we carefully minimize the scalar potential of the theory by including the relevant sub-leading corrections
in order to estimate the deviations of the VEVs from the leading order alignment. The leading order alignment is
not spoiled provided $\langle \varphi_\nu\rangle,\langle \xi\rangle\le \langle \varphi_e\rangle\approx \lambda_c^2 \Lambda$.
In this regime the expected corrections to $\theta_{23}=\pi/4$ and $\theta_{13}=0$ are of order $\lambda_c^2$.
Finally, we extend our construction to the quark sector. The down quark mass hierarchy is reproduced by the breaking of $S_3$,
while the more pronounced hierarchy in the up quark sector requires an additional, spontaneously broken, $Z_3'$ factor in the 
flavour group. The quark mixing angle are correctly reproduced, with the exception of the Cabibbo angle, which requires
an accidental enhancement by a factor 3-4.

\section{A problem with the $\nu_\mu-\nu_\tau$ symmetry}
The most general pattern of lepton masses that automatically leads to $\theta_{13} = \theta_{23}- \pi/4 = 0$ is characterized
by a $\nu_\mu-\nu_\tau$ exchange symmetry, as we briefly recall now.
After breaking of the total lepton number (assumed hereafter) and of the electroweak symmetry, in a two-component spinor notation the lepton mass terms
of the lagrangian read:
\be
{\cal L}=-e^c m_l e-\frac{1}{2}\nu m_\nu \nu
\ee
where $m_l$ and $m_{\nu}$ are the charged lepton mass matrix and the effective neutrino mass matrix.
An arbitrary change of basis in the generation space, realized by means of unitary transformations $\Omega_{e^c}$ and $\Omega_l$ acting on $e^c$ and 
$l=(\nu,e)$ respectively, modifies the form of $m_l$ and $m_{\nu}$, but does not change the physics, in particular
the charged lepton masses $m_e, m_{\mu}, m_{\tau}$, the neutrino masses $m_1,m_2,m_3$ and the mixing matrix $U_{PMNS}$. 
We can exploit this freedom to render diagonal the charged lepton mass matrix $m_l$:
\be \label{flavourbasis}
m'_l= {\rm {diag}} (m_e, m_{\mu}, m_{\tau}), \qquad m'_{\nu}=U_{PMNS}^* {\rm{diag}} (m_1, m_2, m_3)U^\dagger_{PMNS}~~.
\ee
In this basis, the effective neutrino mass matrix is completely determined by the measurable quantities $m_i$ and $U_{PMNS}$. 
Indeed, any complex symmetric $3 \times 3$ matrix has 9 real parameters as the number of parameters in $m_i$ and $U_{PMNS}$. 
In the limit where both $\theta_{13}$ and $\theta_{23}- \pi/4$ vanish, the PMNS mixing matrix becomes (apart from sign convention redefinitions 
and majorana phases):
\begin{equation}  
U_{PMNS}= 
\left(\bad 
 c_{12}&s_{12}&0 \\
   -s_{12}/\sqrt{2}&c_{12}/\sqrt{2}&-1/\sqrt{2}\\
-s_{12}/\sqrt{2}&c_{12}/\sqrt{2}&1/\sqrt{2} \ea
\right) ~,
\label{Unu1}
\end{equation} 
where we are left with the only dependence on the angle $\theta_{12}~$, through $c_{12}\equiv\cos\theta_{12}$ and $s_{12}\equiv\sin\theta_{12}$. 

In the preferred basis of eq. (\ref{flavourbasis}), the conditions  $\theta_{13}=\theta_{23}- \pi/4=0$ correspond to
the requirement that the effective neutrino mass matrix $m_{\nu}$ has the form:
\be \label{allproposal}
m'_{\nu} =  \left( \bad 
c-2b & d & d \\[0.2cm] 
d & b+a & b-a  \\[0.2cm] 
d & b-a & b+a \\ 
               \ea   \right)~,
\ee             
where $a,b,c,d$ are functions of $m_1,m_2,m_3, \theta_{12}~$:
\bea
a&=& 1/2 ~m_3 \nn \\
b&=& 1/2 ~(m_1s^2_{12}+m_2 c^2_{12})\nn \\
c&=& m_1+m_2\nn \\
d&=& 1/\sqrt{2} ~(m_2-m_1)c_{12}s_{12} ~~~.
\eea
All equivalent patterns of lepton mass matrices giving rise to $\theta_{13}=\theta_{23}- \pi/4=0$, are obtained from $m_l'$ diagonal and
$m_\nu'$ of eq. (\ref{allproposal}), by means of $U(3)_{e^c}\times U(3)_l$ transformations.
However, in the flavour basis, where eq. (\ref{flavourbasis}) holds, it is easier to verify that
$m_\nu'$ exhibits an exact $\nu_\mu-\nu_\tau$ exchange symmetry. 
This $\nu_\mu-\nu_\tau$ symmetry
cannot be naively extended to the whole lepton sector. Since $\num$ and $\nut$ are members of the $SU(2)_L$ doublets $(\num,\mu)$
and $(\nut,\tau)$, in such a naive extension the theory would also be symmetric under the exchange of the left-handed charged leptons
$\mu$, $\tau$. But this property is difficult to reconcile with the large mass hierarchy $m_\mu\ll m_\tau$.
 
In some proposals, several Higgs doublets and abelian discrete symmetries of $Z_N$ type are 
introduced to keep diagonal the charged leptons \cite{breaking1, breaking2, lepto, others1},
but the mass hierarchies are obtained by fine-tuning the Yukawa parameters. Alternatively,
we might try to exploit an abelian U(1)$_F$ flavour symmetry to reproduce the observed smallness of $m_\mu/m_\tau$. The simplest
possible charge assignment compatible with the $\nu_\mu-\nu_\tau$ parity is to give equal charges to the left-handed doublets
$(\num,\mu)$ and $(\nut,\tau)$, and different charges to $\mu^c$ and $\tau^c$. However such a choice implies 
a large contribution to $\theta_{23}$ from the charged lepton sector, which would cause large deviations from $\theta_{23}=\pi/4$.

A partial solution to this problem is given by Grimus et al. \cite{Grimus} in a model based on a
flavour symmetry $S_3\times ...$, where dots denotes additional factors of the flavour group.
We shortly review the main features of this model, by adapting it in order to make easier the
comparison with our proposal. Left-handed lepton doublets are chosen as $(1+2)$ of $S_3$: 
$l_1=(\nu_e,e)$ is the invariant singlet and $D_l=((\nu_\mu,\mu),(\nu_\tau,\tau))$ is the $S_3$ doublet 
\footnote{Basic properties of $S_3$ are reviewed in the next section.}. 
Right-handed charged leptons $e^c, \mu^c, \tau^c$ are in a $(1+2^*)$ representation:
the singlet is $e^c$ and the conjugate doublet is $D^c_l=(\mu^c, \tau^c)$. 
The neutrino sector explicitly possesses the $\nu_\mu-\nu_\tau$ symmetry, subgroup
of $S_3$. In the charged lepton sector, three scalar field $\varphi_i$ ($i=1,2,3$) are introduced:
$\varphi_{1,2}$ are singlets of $S_3$ and $\varphi_{3}$ belongs to $1'$.
Thanks to an additional $Z_2$ symmetry under which only $\varphi_1$ and $e^c$ transform non-trivially,
$\varphi_1$ couples with $e^c$ while $\varphi_2$ doesn't. 
The Yukawa interactions of the charged lepton sector are given by
\begin{equation}
{\cal L}_{l}= \frac{h_d}{\Lambda} [ f_1e^c l_1\varphi_1 + f_2(D^c_lD_l)\varphi_2+f_3(D^c_lD_l)'\varphi_3+h.c.]+\ldots
\end{equation}
When the fields $\varphi_1$, $\varphi_2$ and $\varphi_3$ acquire VEVs $v_1$, $v_2$ and $v_3$
respectively, after electroweak symmetry breaking, the charged leptons acquire a diagonal mass matrix
with 
   \begin{eqnarray}
m_e &=& f_1 v_1\nonumber \\
m_{\mu} &=& f_2 v_2+ f_3 v_3 \nonumber \\
m_{\tau} &=& f_2 v_2- f_3 v_3\nonumber
\end{eqnarray}
We see that the non trivial singlet $\varphi_3$ breaks the $\mu-\tau$ symmetry in the charged lepton sector.
In order to obtain the hierarchy between the masses of $\mu $ and $\tau$ a further symmetry, K,  is needed:
\begin{equation}
K: \qquad (\mu^c,\tau^c) \rightarrow (-\mu^c,\tau^c) \,, \qquad \varphi_2 \leftrightarrow \varphi_3
\end{equation}
The K-symmetry imposes $f_2=-f_3$ and $v_2=v_3$ and, when softly broken by $(v_2-v_3)\ll (v_2+v_3)$,
we get:
\begin{equation}
\frac{m_{\mu}}{m_{\tau}}= \frac{v_2-v_3}{v_2+v_3} \ll 1\,.
\end{equation}

The K-symmetry does not commute with $S_3$,
this means that the full symmetry group is not simply  $S_3 \times K$, but is rather generated by
the elements in $S_3$ and in $K$ and consequently much larger than $S_3$.
Furthermore, the smallness of $m_e$ is still not explained by this mechanism.
In our construction, the problem discussed here is completely solved and 
the whole hierarchy $m_e \ll m_{\mu} \ll m_{\tau}$ can be obtained without fine-tuning
and without the need of any ad-hoc symmetry.

\section{A model based on the group $S_3$}

We are thus naturally led to the possibility that the $\nu_\mu-\nu_\tau$ symmetry is only a feature
of the neutrino sector and it is strongly broken in the charged lepton sector.
It is by now well-known how to realize such an hybrid symmetry pattern. 
We embed the $Z_2$ parity arising from the $\nu_\mu-\nu_\tau$ exchange in a larger group $G$. 
Then we selectively couple charged leptons and neutrinos to two different scalar multiplets of $G$, $\varphi_e$ and $(\varphi_\nu,\xi)$,
respectively. Crossed interaction terms of the type $\varphi_e \nu\nu$ and $(\varphi_\nu,\xi) e^c e$ are forbidden by a $Z_3$ 
subgroup of $G$ which will remain unbroken. The multiplets $\varphi_e$ and $\varphi_\nu$ acquire VEVs that break $G$ into 
different subgroups. The VEV of $\varphi_\nu$ breaks $G$ down to $Z_2$, which becomes the residual symmetry in the neutrino sector,
while the VEV of $\varphi_e$ breaks $G$ down to a different subgroup, not containing $Z_2$ and guaranteeing a hierarchical,
quasi diagonal matrix $m_l$.
Here we will make such a construction explicit by taking $G=S_3\times Z_3$. As we will see the powerful of such a broken flavour symmetry is that
it will also produce a hierarchical structure in the charged lepton mass matrices without the need of an additional abelian symmetry.

$S_3$ is group of permutations of three distinct objects and is the smallest non-abelian symmetry group. Geometrically, $S_3$ 
consists of the rotations in three dimensions leaving invariant an equilateral triangle. It has six elements divided into three
conjugate classes: the identity, cyclic and anti-cyclic circulations of triangle apices and
 three exchanges of two of three apices
leaving the third fixed.  Therefore $S_3$ contains three irreducible
representations, which are all real.
\begin{table}[!ht]
\centering
\begin{tabular}{|c||c|c||ccc|}
\hline
&&&&&\\[-9pt]
Classes & n & h & 1 & $1'$ & 2 \\
&&&&&\\[-9pt]
\hline
&&&&&\\[-9pt]
$C_1$ & 1 & 1 & 1 & 1 & 2 \\[3pt]
&&&&&\\[-9pt]
$C_2$ & 2 & 3 & 1 & 1 & -1 \\ [3pt]
&&&&&\\[-9pt]
$C_3$ & 3 & 2 & 1 & -1 & 0 \\[3pt]
\hline
\end{tabular}
\caption{Character table of $S_3$. $C_i$ with i=1,2,3, are the three conjugate classes
of the group; n the order of a class, i.e. the number of distinct elements contained in a class; h is
the order of an element g in a class, i.e. the smallest integer ($>0$) for which $g^h=1$.}
            \label{characterS3}
\end{table}

The $S_3$ group has a presentation given by the generators $S$ and $T$ which satisfy the following
relations:
\be
S^2=T^3=(ST)^2=1~.
\ee
The elements of $S_3$ can be written as products of generators of the group: $1,S,T, ST, TS, T^2$. 
The even permutations are generated by $T$: $\{1, T, T^2 \}$ and form a $Z_3$ subgroup.
The one-dimensional unitary representations are given by
\bea
&1&  S=1, ~~~~~~ T=1 \nn \\
&1'& S=-1, ~~~~ T=1 \nn
\eea
This means that $S_3$ should be broken down to $Z_3$ by the VEV of a pseudo singlet scalar field.
The two-dimensional unitary representation 2 of $S_3$ is given by:
\be \label{bidimofst}
S = \left( \baz 
0 & 1 \\[0.2cm] 
1 & 0 \\ 
               \ea   \right)
\qquad                
T = \left( \baz 
\om & 0 \\[0.2cm] 
0 & \om^2 \\ 
               \ea   \right) ~.
\ee
$S$ corresponds to an interchange $Z_2$ symmetry of the two components 
of a doublet field. For this reason, $S_3$ can be broken down to a $Z_2$ subgroup by the VEV of a doublet scalar field: $\langle \varphi \rangle =(v,v)$. 

The tensor products involving pseudo-singlets are given 
by $1'  \times 1' =1 $ and $1' \times 2=2$. While the product of two doublets is given by 
$2\times 2=2+1+1'$. From the complex representations of $S$ and $T$ given in Eq.~(\ref{bidimofst}), 
one can explicitly calculate these basic tensor products.
Given two doublets $\psi=(\psi_1,\psi_2)^{\rm t}$ 
and $\varphi=(\varphi_1,\varphi_2)^{\rm t}$, it is easy to see that
\be \label{tensorprod1s3}
\ba
\psi_1\varphi_2+\psi_2\varphi_1 \in 1 \\
\psi_1\varphi_2-\psi_2\varphi_1 \in 1' \\
\ea
\qquad
 \left( 
 \ba
 \psi_2\varphi_2 \\
  \psi_1\varphi_1 \\
  \ea
  \right)
  \in 2
 \ee

 The complex conjugate $\psi^*$ belongs to 
the anti-doublet representation $2^*$ for which the representation matrices 
become $(S^*,T^*)$. Defining
\begin{equation}
  \psi' \,\equiv\, \sigma_1\psi^* \,=\, 
  \begin{pmatrix} \psi_2^* \\ \psi_1^* \end{pmatrix} \nn
\end{equation}
and using $\sigma_1 S^* \sigma_1=S$ and $\sigma_1 T^* \sigma_1=T$
one can show that $\psi' $ transforms as a doublet of $S_3$. Then
from Eq. (\ref{tensorprod1s3}) we have
 \be \label{tensorprod2s3}
\ba
\psi_1^* \varphi_1+\psi_2^* \varphi_2 \in 1 \\
\psi_1^* \varphi_1-\psi_2^* \varphi_2 \in 1' \\ 
\ea
\qquad
 \left( 
 \ba
 \psi_1^* \varphi_2 \\
  \psi_2^* \varphi_1 \\
  \ea
  \right)
  \in 2 ~~.
 \ee
The form of tensor product, and consequently the resulting fermion mass pattern,
 depends on the group basis. Other bases are also adopted in the literature. 
 The most important example is the non irreducible permutation basis in which
 the representation matrices for $S_3$ are given by the label-changing matrices.
A study of the various basis of $S_3$ (observe that by a change of basis, it is possible
to make both S and T real) and general forms of mass matrices in models based on $S_3$
can be found in \cite{generals3}.
 
Now we describe a model based on a spontaneously broken $S_3$ flavour symmetry
in which the lepton flavour basis is approximately reproduced by a particular VEV alignment.
To achieve the desired alignment in a simple way, we work with a supersymmetric model,
with N=1 SUSY, eventually broken by small soft breaking terms.
The left-handed doublets transform as $(1+2)$ of $S_3$ and we will call 
$l_1=(\nu_e,e)$ the invariant singlet and $D_l=((\nu_\mu,\mu),(\nu_\tau,\tau))$ the $S_3$ doublet. 
The right-handed charged leptons $e^c, \mu^c, \tau^c$
are all in the non-trivial singlet representation $1'$. The flavour symmetry is broken by two 
doublets $\varphi_e, \varphi_\nu$ and a singlet $\xi$. We assume that the flavon fields which break
the $S_3$ symmetry are gauge singlets. In addition, we introduce an extra abelian symmetry
$Z_3$ in such a way that  $\varphi_\nu$ and $\xi$ couple only to the neutrino sector and
$\varphi_e$ to the charged lepton sector, at the leading order.
We summarize the transformation rules of the fields in Table~\ref{chargesS3}.
\begin{table}[!ht] 
\centering
\begin{tabular}{|c||c|c|c|c|c|c||c|c|c|}
\hline
&&&&&&&&&\\[-9pt]
Field & $h_{u,d}$ & $l_1$ & $D_l$ & $e^c$ & $\mu^c$ & $\tau^c$ & $\varphi_e$ & $\varphi_\nu$ & $\xi$ \\
&&&&&&&&&\\[-9pt]
\hline
$S_3$ &1 & 1 & 2 & $1'$ & $1'$ & $1'$ & 2 & 2 & 1 \\[3pt]
\hline
$Z_3$ & 1 &  $\omega$ & $\omega$ & $\omega^2$ & $\omega^2$ & $\omega^2$ &1 & $\omega$ & $\omega$ \\ [3pt]
\hline
\end{tabular}
\caption{Transformation properties of matter and flavon fields under the flavour group.}
\label{chargesS3}
\end{table}

The Yukawa interactions for the lepton sector are controlled by the superpotential
\be
w=w_\nu+w_e
\label{ww}
\ee
where we have separated the contribution to neutrino masses and to charged lepton masses. 
For neutrinos we have:
\be
w_\nu= \frac{h_u^2}{\Lambda^2} (y_1D_lD_l\varphi_\nu+y_2D_lD_l\xi+2 y_3l_1D_l\varphi_\nu+
y_4l_1l_1\xi)+\ldots
\label{wwnu}
\ee
where dots stand for higher-order contributions.
As we will see later, $\varphi_\nu$ and $\xi$ develop VEVs of the type
\be
\langle \varphi_\nu \rangle= (v_\nu,v_\nu)~~~,~~~~~~~\langle \xi \rangle= u~~~.
\label{vevsnu}
\ee
After the flavour symmetry breaking and the electroweak symmetry breaking, $w_\nu$ becomes:
  \bea
w_\nu&=& \frac{h_u^2 }{\Lambda^2} v_\nu (y_1 \num \num+y_1 \nut \nut +2 y_3 \nu_e \num + 2 y_3\nu_e \nut) \nn\\
&+&  \frac{h_u^2 }{\Lambda^2} u(2 y_2 \num\nut+ y_4\nu_e\nu_e)
+\ldots \nn
\eea
giving rise to the following neutrino mass matrix:
\be
m_{\nu} =  \frac{2 h_u^2}{\Lambda^2}v_\nu
\left( 
\bad 
y_4 x & y_3 & y_3 \\[0.2cm] 
y_3  & y_1& y_2 x \\[0.2cm] 
y_3 & y_2 x& y_1  \\ 
\ea   
\right)~~~~~~~~~~~~~~~~~~~x\equiv\frac{u}{v_\nu}~~~.
\label{massnu1}
\ee
This mass matrix is of the form (\ref{allproposal}) since $w_\nu$ respects explicitly the 
$\num-\nut$ interchange symmetry. It is diagonalized by sending $\nu$ into $U_\nu \nu$, where
\begin{equation}  
U_{\nu}= 
\left(\bad 
 c_{12}&s_{12}&0 \\
   -s_{12}/\sqrt{2}&c_{12}/\sqrt{2}&-1/\sqrt{2}\\
-s_{12}/\sqrt{2}&c_{12}/\sqrt{2}&1/\sqrt{2} \ea
\right) ~~~~.
\label{Unu2}
\end{equation} 

In the charged lepton sector, the leading terms of the superpotential are given by:
\be
\label{lcharged}
w_e= \gamma  \tau^c (D_l\varphi_e)' ~\frac{h_d}{\Lambda}+
(\beta' \mu^c +\gamma' \tau^c)(D_l\varphi_e\varphi_e)'~ \frac{h_d}{\Lambda^2}+
(\alpha'' e^c +\beta'' \mu^c + \gamma'' \tau^c)~l_1~ (\varphi_e\varphi_e\varphi_e)'~ \frac{h_d}{\Lambda^3}+...
\ee
where dots stand for additional operators of order $1/\Lambda^3$, to be specified in the next section.
The electroweak singlets $e^c$, $\mu^c$ and $\tau^c$ have the same quantum numbers and above, without loosing generality, we have defined
$\tau^c$ as the field coupled to  $(D_l\varphi_e)'$ and $(\beta' \mu^c +\gamma' \tau^c)$ as the combination coupled to
$(D_l\varphi_e\varphi_e)'$. As we will see $\varphi_e$ acquires a VEV of the type:
\be
\langle \varphi_e \rangle= (v,0)~~~.
\label{vevse}
\ee
Such a VEV breaks the parity symmetry generated by $S$ in a maximal way, since 
\be
\langle \varphi_e \rangle^\dagger S \langle \varphi_e \rangle=0~~~.
\ee
We get:
\be
w_e=\gamma  \tau^c \tau v_d \left(\frac{v}{\Lambda}\right)+(\beta' \mu^c +\gamma' \tau^c)~\mu v_d\left(\frac{ v}{\Lambda}\right)^2+
(\alpha'' e^c +\beta'' \mu^c + \gamma'' \tau^c)~e v_d\left( \frac{v}{\Lambda}\right)^3~~~.
\ee
Notice that, provided the ratio $v/\Lambda$ is much smaller than one, $w_e$ generates
a hierarchical mass pattern, as desired: $m_e \ll m_{\mu} \ll m_{\tau}$.
We see that the ratio $v/\Lambda$ should be approximately equal to the ratio $m_{\mu}/m_{\tau}$.
We also notice that $(\varphi_e\varphi_e)'=0$ due to anti-symmetry and consequently the electron mass is not generated 
at the same order as the muon mass{\footnote{Observe that a term $(\alpha'' e^c +\beta'' \mu^c + \gamma'' \tau^c)~l_1~ 
 (\varphi_\nu\varphi_\nu\varphi_\nu)'~ h_d/\Lambda$ would be allowed by $S_3$ and $Z_3$ symmetries
  but  $(\varphi_\nu\varphi_\nu\varphi_\nu)'=0$ due to the specific VEV alignment $\langle \varphi_\nu \rangle
  = v_\nu (1,1)$. Similarly $(\varphi_\nu\varphi_\nu\xi)'=0$}}. 
This is a very nice result because we will end up with a nearly diagonal and
hierarchical $m_l$ without a Froggatt-Nielsen mechanism based on a continuous U(1) flavour symmetry.
 
We define the expansion parameter
\be
v/\Lambda \equiv \lambda^2 \ll 1~~~.
\ee 
The resulting charged lepton mass is of the type 
\be \label{masslleft}
m_{l} =   \left( \bad 
\alpha'' \lambda^4 & 0 & 0 \\[0.2cm] 
\beta'' \lambda^4 & \beta' \lambda^2 & 0  \\[0.2cm] 
\gamma'' \lambda^4 & \gamma' \lambda^2 & \gamma \\ 
               \ea   \right)v_d \lambda^2~,
\ee 
with approximate eigenvalues:
\be
m_\tau\approx \gamma \lambda^2 v_d~~~,~~~~~~~m_\mu\approx \beta' \lambda^4 v_d~~~,~~~~~~~m_e\approx \alpha'' \lambda^6 v_d~~~.
\ee
In order to reproduce correctly the charged lepton hierarchy, $\lambda \approx \lambda_c$ 
where $\lambda_c$ is the Cabibbo angle.
The adimensional constants $\gamma$, $\gamma'$, $\beta'$, $\gamma''$, $\beta''$, $\alpha''$ are 
numbers with absolute value of order one and we find that the transformation needed to diagonalize $m_l$ is:
\be
V_e^T m_l U_e=diag(m_e,m_\mu,m_\tau)~~~,
\ee
where the unitary matrix $U_e$, parametrized in the standard way, involves rotations of order
$\theta_{12}^e=O(\lambda^2)$, $\theta_{13}^e=O(\lambda^4)$, $\theta_{23}^e=O(\lambda^2)$. 
The lepton mixing matrix is then $U_{PMNS}=U_e^\dagger U_{\nu}$ where $U_{\nu}$ is given by Eq.~(\ref{Unu2}).
We see that $U_l$ introduces deviations of $\theta_{13}$ and 
$\theta_{23}$ from $0$ and $\pi/4$, both of order $\lambda^2$:
\be
\theta_{13}=O(\lambda_c^2)~~~,~~~~~~~\theta_{23}=\frac{\pi}{4}+O(\lambda_c^2)~~~,
\label{loresult}
\ee
Additional deviations induced by sub-leading contributions will be discussed in the next section.

\section{Vacuum alignment}

In this section we discuss the minimization of the scalar potential leading to the results given in eqs.
(\ref{vevsnu},\ref{vevse}). 
Notice that the superpotential $w$ of eqs. (\ref{ww},\ref{wwnu},\ref{lcharged}) is also invariant under a continuous $U(1)_R$
symmetry under which matter fields have $R=+1$, while Higgses and flavons have $R=0$. Such a symmetry will be eventually broken
down to R-parity
by small SUSY breaking effects that can be neglected in first approximation in our analysis. The superpotential must have $R=2$ and, to obtain
non-trivial minima for the flavon fields, we introduce two driving multiplets with $R=2$:
a full singlet $\chi$ under the flavour group and a doublet $\psi$, transforming as $(2,\omega)$
under $S_3\times Z_3$. Under these assumptions, at the leading order the superpotential depending on the driving fields
is given by:
\bea
w_d&=&a\chi \varphi_e^2 + b\psi \varphi_\nu^2+c\psi\varphi_\nu \xi\nn\\
&=&2 a \chi {\varphi_e}_1 {\varphi_e}_2+
b \psi_1 ({\varphi_\nu}_1)^2+b\psi_2 ({\varphi_\nu}_2)^2+
c\xi \psi_1 {\varphi_\nu}_2+c\xi \psi_2 {\varphi_\nu}_1~~~.
\label{wd}
\eea
In the SUSY limit the condition for the minima are:
\bea
\frac{\partial w_d}{\partial\chi}&=&2 a{\varphi_e}_1 {\varphi_e}_2=0\nn\\
\frac{\partial w_d}{\partial\psi_1}&=& b ({\varphi_\nu}_1)^2+ c\xi {\varphi_\nu}_2=0\nn\\
\frac{\partial w_d}{\partial\psi_2}&=& b ({\varphi_\nu}_2)^2+ c\xi {\varphi_\nu}_1=0~~~,
\eea
This set of equations admit the solution:
\be
\langle \varphi_e \rangle= (v,0)~~~,~~~~~~~
\langle \varphi_\nu \rangle= -\frac{c}{b} (u,u)~~~,~~~~~~~\langle \xi \rangle= u~~~,
\label{lomin}
\ee
with $v$ and $u$ arbitrary complex numbers. The flat directions along ${\varphi_e}_1$
and $\xi$ can be removed by the interplay of radiative corrections to the scalar potential
and soft SUSY breaking terms. As we have seen, to correctly reproduce the hierarchy of
charged fermion masses we must assume $v/\Lambda=O(\lambda_c^2)$.
The alignment of $\varphi_e$ and $\varphi_\nu$ is modified when higher-dimensional operators 
are included in the analysis. At the next-to-leading order, the superpotential acquires the
additional contribution
\be
\delta w_d=\sum_{i=1}^8 \frac{d_i}{\Lambda} I_i
\ee
where $I_i$ is a basis of independent invariants of dimension four:
\bea
I_1&=&\chi \varphi_e^3\nn\\
I_2&=&\chi \varphi_\nu^3\nn\\
I_3&=&\chi \varphi_\nu^2 \xi\nn\\
I_4&=&\chi \xi^3\nn\\
I_5&=&(\psi\varphi_e)(\varphi_\nu^2)\nn\\
I_6&=&(\psi\varphi_\nu)(\varphi_e\varphi_\nu)\nn\\
I_7&=&\psi\varphi_\nu\varphi_e \xi\nn\\
I_8&=&\psi\varphi_e \xi^2~~~.
\eea
We look for a new SUSY minimum of the scalar potential generated by $w_d+\delta w_d$. By working 
around the leading-order minimum (\ref{lomin}), we find
\be
\langle \varphi_e \rangle= (v,\delta v)~~~,~~~~~~~
\langle \varphi_\nu \rangle= -\frac{c}{b} (u+\delta u_1,u+\delta u_2)~~~,~~~~~~~\langle \xi \rangle= u~~~,
\label{nlomin}
\ee
with $v$ and $u$ still arbitrary and
\bea
\frac{\delta v}{v}&=&
\left[-\frac{d_1}{2a}+\frac{(2 c^3 d_2-2 b c^2 d_3-b^3 d_4)}{2 a b^3}\left(\frac{u}{v}\right)^3\right]\frac{v}{\Lambda}\nn\\
\frac{\delta u_1}{u}&=&\left[-\frac{c^2(2d_5+3d_6)-2 b c d_7+b^2 d_8}{3 b c^2}\right] \frac{v}{\Lambda}\nn\\
\frac{\delta u_2}{u}&=&\left[-\frac{c^2(4d_5+3d_6)-2 b c d_7+2 b^2 d_8}{3 b c^2}\right] \frac{v}{\Lambda}~~~.
\eea
We see that $\delta u_{1,2}/u$ are of order $v/\Lambda\approx \lambda_c^2$ and, barring accidental relations, they are different,
thus modifying the leading order alignment $\varphi_\nu\propto (1,1)$. We see that also $\delta v/v$ is of order $v/\Lambda\approx \lambda_c^2$,
provided $u\le v$. This last requirement is important in order to maintain the pattern of $m_l $ discussed above.

The mass matrices $m_\nu$, eq. (\ref{massnu1}), and $m_l$, eq. (\ref{masslleft}), are modified by subleading terms in the expansion
in $1/\Lambda$. The neutrino mass matrix is modified by terms of relative order $\lambda_c^2$ by two independent effects:
terms originating from higher-order operators containing an additional insertion of $\varphi_e/\Lambda$ and terms
originating from the modification to the leading order vacuum.

Coming to the charged lepton mass matrix, the higher dimensional operators whose effect cannot be absorbed in the 
parameters $\gamma$, $\gamma'$, $\beta'$, $\gamma''$, $\beta''$, $\alpha''$ are those contributing to the elements
12, 13 and 23, that vanish at the leading order. They are:
\be
f^c D_l \varphi_\nu^3\frac{h_d}{\Lambda^3}~~~,~~~~~~~f^c D_l \varphi_\nu^2 \xi\frac{h_d}{\Lambda^3}~~~,
~~~~~~~f^c D_l \varphi_\nu \xi^2\frac{h_d}{\Lambda^3}
\ee
where $f^c=(e^c,\mu^c,\tau^c)$. Their contribution is of order $(u/\Lambda)^3\le \lambda_c^6$ and they do not spoil our conclusion
about the contribution $U_e$ of $m_l$ to the mixing matrix $U_{PMNS}$. Finally we should account for the modified vacuum, eq. (\ref{nlomin}).
We find that also this effect does not modify the order of magnitude of the $U_e$ mixing matrix. 
In conclusion, there are no further contribution to the lepton mixing coming from $m_l$ and the modification
to $m_\nu$ are of relative order $\lambda_c^2$. Therefore our previous results, eq. (\ref{loresult}),
are stable under inclusion of subleading effects.

\section{Quark masses and mixing angles}

The flavour group $S_3\times Z_3$, with few additional ingredients, can also provide a satisfactory description of the quark sector.
We recall three special features of quarks.
First, the top mass is very close to the electroweak symmetry breaking scale, $v= \sqrt{v_u^2+v_d^2} \approx 174 ~\rm{GeV} $,
indicating an unsuppressed, renormalizable Yukawa coupling for the heaviest quark. 
Second, the hierarchy among quark masses of the up type is 
much more pronounced than the hierarchy in the down quark sector. Third, the mixing angle $\lambda_c$ 
between the first two generations is the dominant one. 
The first feature suggests to adopt for quarks a different $S_3$ assignment. 
We still put left-handed quarks in a $(1+2)$ representation of $S_3$, as for the left-handed leptons.
We call $q_3=(t,b)$ the invariant singlet and $D_q=((u,d),(c,s))$ the $S_3$ doublet.
At variance with the lepton sector, we assign all right-handed quarks to the invariant singlet $1${\footnote{Observe that there is 
another equivalent choice of classification scheme under $S_3$ for quarks:  
$q_i$ transform as $(1'+2)$ and the right-handed quarks transform as $1'$.
The resulting quark mass matrices remain unchanged.}}. 
As a consequence, there is one combination of right-handed quarks of up-type, denoted by $t^c$, directly coupled to a left-handed
quark $q_3$, the invariant singlet under $S_3$, thus providing a renormalizable Yukawa interaction for the top quark.
The second feature requires an extension of the flavour group, to differentiate up and down quarks.
A simple extension is provided by an additional $Z_3'$ factor. The flavour symmetry becomes $S_3\times Z_3\times Z_3'$.
Leptons do not transform under the new $Z_3'$, so that the whole construction of the previous sections remains unchanged.
In the quark sector we assume the transformation properties:
\bea
u^c&\to& \omega~ u^c\nn\\
(d^c,s^c,b^c,c^c)&\to& \omega^2~ (d^c,s^c,b^c,c^c)~~~,
\eea
and the remaining fields are taken invariant under $Z_3'$.
To complete the construction we introduce a new flavon field, $\xi'$, transforming only under $Z_3'$ as $\xi'\to \omega \xi'$,
and developing a large VEV:
\be
\langle\xi'\rangle=u'~~~,~~~~~~~\frac{u'}{\Lambda}\equiv {\lambda'}^2.
\ee
These transformation properties are summarized in table 2.
\begin{table}[!ht] 
\centering
\begin{tabular}{|c||c|c|c|c|c|c||c|}
\hline
&&&&&&&\\[-9pt]
Field & $D_q$ & $q_3$ & $d^c,~s^c,~b^c$ & $u^c$ & $c^c$ & $t^c$ & $\xi'$ \\
&&&&&&&\\[-9pt]
\hline
$S_3$ &$2$ & $1$ & $1$ & $1$ & $1$ & $1$ & $1$ \\[3pt]
\hline
$Z_3$ & $\omega$ &  $\omega$ & $\omega^2$ & $\omega^2$ & $\omega^2$ & $\omega^2$ & $1$ \\ [3pt]
\hline
$Z_3'$ & $1$ &  $1$ & $\omega^2$ & $\omega$ & $\omega^2$ & $1$ &$\omega$ \\ [3pt]
\hline
\end{tabular}
\caption{Transformation properties of quarks and the flavon field $\xi'$ under the flavour group.}
\label{chargesq}
\end{table}
With this charge assignment, the leading operators contributing to quark masses are:
\be
w_q=w_{down}+w_{up}~~~.
\ee
In the down sector we have:
\be
w_{down}= \gamma_d ~b^c q_3~\frac{\xi'}{\Lambda} h_d+
(\beta_d' s^c +\gamma_d' b^c)~(D_q\frac{\varphi_e}{\Lambda})~\frac{\xi'}{\Lambda} h_d+
(\alpha_d'' d^c +\beta_d'' s^c + \gamma_d'' b^c)~ (D_q\frac{\varphi_e\varphi_e}{\Lambda^2})~\frac{\xi'}{\Lambda} h_d+...
\ee
where we have defined
$b^c$ as the field coupled to  $q_3$ and $(\beta_d' s^c +\gamma_d' b^c)$ as the combination coupled to
$(D_q\varphi_e)$. For up type quarks we obtain:
\bea
w_{up}&=& \left(\alpha_u \frac{{\xi'}^2}{\Lambda^2}~u^c+\beta_u \frac{\xi'}{\Lambda}~ c^c+\gamma_u t^c\right) q_3 h_u\nn\\
&+&\left(\alpha_u' \frac{{\xi'}^2}{\Lambda^2}~u^c+\beta_u'\frac{\xi'}{\Lambda}~ c^c +\gamma_u' t^c\right)(D_q\frac{\varphi_e}{\Lambda})~ h_d\nn\\
&+&\left(\alpha_u''\frac{{\xi'}^2}{\Lambda^2}~u^c+\beta_u'' \frac{\xi'}{\Lambda}~ c^c + \gamma_u'' t^c\right)~ (D_q\frac{\varphi_e\varphi_e}{\Lambda^2})~ h_d+...
\eea
The superpotential $w_q$, after electroweak and flavour symmetry breaking, gives rise to the following mass terms:
\bea
w_q&=&\gamma_d b^c~ b~ v_d{\lambda'}^2+
(\beta_d' s^c +\gamma_d' b^c)~s~ v_d\lambda^2 {\lambda'}^2+
(\alpha_d'' d^c +\beta_d'' s^c + \gamma_d'' b^c)~d~ v_d{\lambda^4} {\lambda'}^2\nn\\
&+& \left(\alpha_u {\lambda'}^4~u^c+\beta_u {\lambda'}^2~ c^c+\gamma_u t^c\right) t~ v_u\nn\\
&+&\left(\alpha_u' {\lambda'}^4~u^c+\beta_u' {\lambda'}^2~ c^c +\gamma_u' t^c\right) c~ v_u \lambda^2\nn\\
&+&\left(\alpha_u''{\lambda'}^4~u^c+\beta_u'' {\lambda'}^2~ c^c + \gamma_u'' t^c\right)~ u~ v_u \lambda^4+...
\eea
Assuming that the VEV of $\xi'$ is similar to the VEV of $\xi$ and $\varphi_e$:
\be
\frac{u'}{\Lambda}\equiv {\lambda'}^2\approx \frac{v}{\Lambda}\equiv {\lambda}^2~~~,
\ee
we find that the quark mass matrices have the following pattern
\be \label{massdownleft}
m_{d} =   \left( \bad 
\lambda^4 & 0 & 0 \\[0.2cm] 
\lambda^4 & \lambda^2 & 0  \\[0.2cm] 
\lambda^4 & \lambda^2 & 1 \\ 
               \ea   \right) v_d\lambda^2~.
\ee

 \be \label{massupleft}
m_{u} =   \left( \bad 
\lambda^8 & \lambda^6 & \lambda^4 \\[0.2cm] 
\lambda^6 & \lambda^4 & \lambda^2  \\[0.2cm] 
\lambda^4 & \lambda^2 & 1 \\ 
               \ea   \right) v_u~~~,
\ee
where we have set to one the unknown coefficients $\alpha_{u,d},...$
We see that the quark mass hierarchy is correctly reproduced by the interplay of the 
symmetry breaking parameters $\lambda$ and $\lambda'$, once we take $\lambda\approx\lambda'\approx\lambda_c$.
\bea
m_t : m_c : m_u &\approx& 1 : \lambda_c^4 : \lambda_c^8 \nn\\
m_b : m_s : m_d &\approx& 1 : \lambda_c^2 : \lambda_c^4 \nn ~~~.
\eea
We have sufficiently many order-one free parameters $\alpha_{u,d},...$ to provide a good fit to
the quark masses. Also the $V_{CKM}$ mixing matrix is well reproduced,
at least at the level of the powers of $\lambda_c$, with the only exception of the Cabibbo angle,
which in this model is a combination of two independent contributions of order $\lambda_c^2$.
We need an accidental enhancement of this combination in order to obtain the correct
Cabibbo angle.

In summary, our flavour symmetry can be easily extended to the quark sector.
Almost all quark masses and quark mixing angles are satisfactorily described by the spontaneous breaking of the
symmetry $S_3\times Z_3\times Z_3'$. All dimensionless coefficients of the lagrangian are numbers of order one.
No ad-hoc relations are required, with the exception of a moderate tuning needed to reproduce the Cabibbo angle.
Unfortunately, the number of free parameters, the order-one coefficients $\alpha_{u,d},...$, is too big to
allow to formulate a quantitative, testable prediction, beyond the order-of-magnitude estimates illustrated above.
A generic prediction of our model is that $\tan\beta=v_u/v_d$ is of the order one, since the hierarchy
between top and bottom quark is ascribed to a symmetry breaking parameter, ${\lambda'}^2$, rather than to
$v_u\gg v_d$.

\section{Conclusion}
In the near future significant improvements on the parameters of the lepton mixing matrix are expected. 
The angles $\theta_{13}$ and $\theta_{23}$ will be constrained with an accuracy of about $\lambda_c^2$, 
and it will be possible to establish more precisely how close are $\theta_{13}$ and $\theta_{23}$ to zero and to $\pi/4$, respectively. 
This will allow to discriminate between different models of fermion masses. Many models predict a generically large $\theta_{23}$ 
and a generically small $\theta_{13}$, with deviations of order one from the reference values $\theta_{13}=0$ and $\theta_{23}=\pi/4$. 
Only in a selected subset of the existing models a nearly maximal $\theta_{23}$ and a nearly vanishing $\theta_{13}$ are expected.
Even in these last `special' models $\theta_{13}=0$ and $\theta_{23}=\pi/4$ generally arise only as leading order results of 
a power series expansion. The expansion parameter is the ratio $\langle \varphi \rangle/\Lambda$ between the VEV of some field spontaneously 
breaking the underlying flavour symmetry, and the cut-off $\Lambda$ of the theory. 
Therefore it is of great interest to provide an accurate estimate of the expected deviations from the leading order predictions.
Here we have analyzed one of the simplest models that enforce $\theta_{13}\approx 0$ and $\theta_{23}\approx\pi/4$,
thanks to an approximate $\nu_\mu-\nu_\tau$ parity symmetry of the neutrino sector. This parity symmetry is part of a larger
flavour symmetry, $S_3\times Z_3$, of the whole lepton sector, spontaneously broken along two different directions
for neutrinos and charged leptons, respectively, as a consequence of a specific vacuum alignment. 
In the charged lepton sector the $\nu_\mu-\nu_\tau$ parity symmetry is maximally
broken. A noticeable feature of the model is that the mass hierarchies between charged leptons are completely 
determined by the symmetry breaking of the permutation group, without the need of any additional ad-hoc ingredient.  
We find $\theta_{23}=\pi/4+O(\lambda_c^2)$ and $\theta_{13}=O(\lambda_c^2)$, where the
$O(\lambda_c^2)$ corrections come from the contribution of the charged lepton sector and are related to the mass
hierarchies. We have carefully analyzed subleading effects due to higher dimensional operators that modify the Yukawa couplings and the 
vacuum alignment and we have verified that they do not spoil the above predictions. 
We have also briefly discussed how to extend the model to the quark sector,
where, with a minimal enlargement of the flavour symmetry, masses and mixing angles are correctly reproduced,
with the exception of the mixing angle between the first two generations, that requires a small accidental enhancement.
\vskip 0.5 cm
\section*{Acknowledgements}
We thank Guido Altarelli for useful discussion. We recognize that this work has been partly supported by 
the European Commission under contracts MRTN-CT-2004-503369 and MRTN-CT-2006-035505.

\end{document}